\documentclass[a4paper,11pt]{article}
\usepackage{jheppub} 
\usepackage{lineno}

\arxivnumber{2605.20791}

\title{\boldmath First mass determination of electroweak vortex rings in the Standard Model}
\author[a,b]{Dan Zhu,}
\author[a,b]{Qingyue Zhang,}
\author[c]{Khai-Ming Wong}
\author[a,b]{and Xurong Chen}
\affiliation[a]{Institute of Modern Physics,\\
	XinQiaoBeiLu 1, Huizhou 516000, China}
\affiliation[b]{University of Chinese Academy of Sciences,\\
	YuQuanLu 19A, Beijing 100049, China}
\affiliation[c]{Universiti Sains Malaysia,\\
	Jalan Sungai Dua, Penang 11800 USM, Malaysia}

\emailAdd{xchen@impcas.ac.cn}

\abstract{We report the first rigorous evaluation of the physical mass of electroweak vortex rings, establishing precise values of 18.01 and 26.80~TeV for solutions characterized by different winding numbers. Analysis of the internal structure reveals that repulsive interactions shape the geometry of these configurations, while complex current distributions lead to a neutral analogue of Amp\`ere’s circuital law, suggesting a corresponding self-stabilizing pinch mechanism. These findings set the energy scales for the potential observation of such configurations at future colliders and offer a framework for understanding topological structures in the Standard Model.}

\begin{document}
	\maketitle
	\flushbottom

	\section{Introduction}
	Topological defects, such as monopoles and cosmic strings, arise naturally from the spontaneous symmetry breaking of fundamental gauge symmetries and have profoundly influenced diverse disciplines ranging from cosmology to condensed matter physics~\cite{Volovik}. These structures have been widely studied in various gauge theories and models, with their internal architectures becoming particularly rich under axial symmetry~\cite{RaduVolkov}. Among these, vortex rings~\cite{ymhvortex,Amin,wsvortex1,wsvortex2,Teh}---closed flux tubes of gauge fields---have attracted broad interest due to their far-reaching implications in superfluids and superconductors~\cite{v1,v2,v3,v4,v5,v6,v7,v8}.
	
	The study of vortex rings in the context of non-Abelian gauge theories has a history spanning decades. Early numerical breakthroughs were achieved in the SU(2) Yang-Mills-Higgs (YMH) theory, where axially symmetric vortex ring solutions were first constructed by Kleihaus, Kunz, and Shnir~\cite{ymhvortex}. These configurations can be interpreted as the topological structures formed by a merging monopole-antimonopole pair (MAP)~\cite{KKMAP} when the winding number exceeds three; evidence for such an MAP-to-vortex transition can be found in refs.~\cite{Amin,KKTrans}. Subsequently, the idea was extended to the Weinberg-Salam (electroweak) model, first in the limit of a vanishing weak mixing angle $\theta_W \to 0$ by Kleihaus, Kunz, and Leissner~\cite{wsvortex1,wsvortex2}, where the U(1) gauge potential decouples and the configuration effectively reduces to an SU(2) vortex ring embedded in the electroweak theory. The first attempt to include a nonzero Weinberg angle was made by Teh, Ng, and Wong~\cite{Teh}, who obtained numerical solutions for a simplified case where $\theta_W=\pi/4$. However, the solution profiles are not completely smooth and the convergence suffered greatly. Further study of this configuration has been hindered by severe numerical challenges.
	
	In this work, we present high-precision numerical solutions to the electroweak vortex rings with $\theta_W$ treated as a free parameter, enabling a direct calculation of the total energy. This provides, to our knowledge, the first precise determination of the mass of topological structures within the Standard Model, distinct from prior estimates that relied on simplified limits or model extensions~\cite{Re1,Re2,Re3,Re4,Re5,Re6,Re7}. It is therefore important to clarify the physical nature of the solutions we seek. These configurations are static, saddle-point solutions to the electroweak field equations. In contrast to the SU(2) YMH theory, the electroweak vacuum manifold is $\mathcal{M}_\text{EW}\simeq S^3$, for which the second homotopy group is trivial. Hence, these configurations do not possess a conserved topological monopole charge and are not protected against decay by $\pi_2(\mathcal{M}_\text{EW})$. Moreover, the nontrivial third homotopy group, $\pi_3(\mathcal{M}_\text{EW})=\mathbb{Z}$, implies the existence of non-contractible loops in the electroweak configuration space, which are associated with energy barriers between neighboring classical vacua~\cite{Manton}. The solutions we seek are located at the top of these barriers, constitute saddle points of the energy functional, and are therefore unstable. At finite temperature, thermal fluctuations may overcome these barriers, leading to baryon-number-violating processes~\cite{wsvortex2}. This stems from the fact that these configurations carry half-integer Chern-Simons number, and chiral anomalies relate changes in Chern-Simons number to the generation of baryon number. We now summarize our main contributions, which are threefold.
	
	First, the stabilization paradigm recently established for Cho-Maison MAPs~\cite{ChoMaison,ZhuDan2,ZhuDan3}---repulsive interactions mediated by the Higgs and $Z$-bosons counterbalancing attractive forces---is confirmed to govern the electroweak vortex rings. A crucial distinction is that these configurations carry no magnetic charge, confirming that the repulsive interactions are an intrinsic and fundamental feature of the Weinberg-Salam model, not contingent upon the presence of a monopole. We show how the interplay of these forces determines the geometric size (radius $R_\rho$) and the total energy $E$ of the vortex rings. For physical parameters, we obtain $E_{(n=2)} \approx 18.01$ TeV and $E_{(n=3)} \approx 26.80$ TeV, setting the scales for future collider searches. These findings lead to the conclusion that their role in electroweak baryogenesis is expected to be subdominant: their energies (18--27 TeV) are significantly larger than the ordinary sphaleron barrier ($\sim10$ TeV), leading to a stronger Boltzmann suppression in the thermal plasma.
	
	Second, we provide a comprehensive visualization of the vortex rings' internal structures. The electromagnetic and $Z$-boson currents are both steady, azimuthal flows, with their respective field lines forming concentric loops in the meridian plane. This structure is suggestive of a neutral analogue of the classical Amp\`ere's circuital law, where a circulating neutral ``electric" current generates a concentric neutral ``magnetic" field, a direct consequence of the electroweak field equations that link the curl of the non-Abelian magnetic field to the Higgs current. In contrast, the charged $W$-boson currents are far more intricate, forming two parallel tori whose $(\hat{r}_i,\hat\theta_i)$ and $\hat\phi_i$ components are in quadrature, reminiscent of a toroidal-poloidal knot-like structure.
	
	Third, through a stress-energy tensor analysis, an intense pressure spike is observed at precisely the vortex ring locations, representing an attractive ``pinch" generated by the circulating currents. This pinching effect is the logical corollary to the aforementioned neutral analogue of Amp\`ere's law, akin to the electromagnetic pinch in plasma physics but mediated by the $Z$-boson. Furthermore, when integrated, these components exhibit pronounced non-monotonic, oscillatory behaviors as functions of both the Higgs self-coupling and the Weinberg angle. This signals a complex dynamical evolution that could have interesting implications for the stability and possible decay channels of these configurations in the early universe.

	\section{Model and ansatz}
	\subsection{Bosonic sector of the Weinberg-Salam model}
	To establish the theoretical basis for electroweak vortex rings, we summarize the bosonic sector of the Weinberg-Salam model and its equations of motion. This model is described by the following Lagrangian density~\cite{wsvortex1,wsvortex2,Teh}
	\begin{equation}
		\mathcal{L}=-\frac{1}{4}F^a_{\mu\nu}F^{a\mu\nu}-\frac{1}{4}f_{\mu\nu}f^{\mu\nu}-\left(\mathcal{D}_\mu\boldsymbol{\phi}\right)^\dagger\left(\mathcal{D}^\mu\boldsymbol{\phi}\right)-\frac{\lambda}{2}\left(\boldsymbol{\phi}^\dagger\boldsymbol{\phi}-v^2/2\right)^2,\label{eqn:Lagrangian}
	\end{equation}
	where $\boldsymbol\phi$ is the complex scalar Higgs doublet, $\lambda$ parameterizes the Higgs self-coupling strength, and $v$ denotes the vacuum expectation value. The electroweak covariant derivative is defined as
	\begin{equation}
		\mathcal{D}_\mu\equiv \partial_\mu-\frac{\mathrm{i}}{2}gA^a_\mu\sigma^a-\frac{\mathrm{i}}{2}g'B_\mu,\label{eqn:covariant_derivative}
	\end{equation}
	where $\sigma^a$ stands for the Pauli matrices, $A^a_\mu,\,B_\mu$ are the SU(2) and U(1) gauge potentials, with their respective coupling constants $g$ and $g'$. Finally, the corresponding field strength tensors are
	\begin{align}
		F^a_{\mu\nu}&=\partial_\mu A^a_\nu-\partial_\nu A^a_\mu+g\varepsilon^{abc}A^b_\mu A^c_\nu,\\
		f_{\mu\nu}&=\partial_\mu B_\nu-\partial_\nu B_\mu.
	\end{align}
	
	Variations of the action corresponding to eq.~\eqref{eqn:Lagrangian} yield the stress-energy tensor and the equations of motion:
	\begin{align}
		T_{\mu\nu}&=2(\mathcal{D}_\mu\boldsymbol{\phi})^\dagger(\mathcal{D}_\nu\boldsymbol{\phi})+F_\mu^{a\beta}F_{\nu\beta}^a+f_\mu^\beta f_{\nu\beta}+g_{\mu\nu}\mathcal{L},
		\label{eqn:T}\\
		D^\mu F^a_{\mu\nu}&=gJ^a_\nu=\frac{\mathrm{i}g}{2}\left[\boldsymbol{\phi}^\dagger\sigma^a\left(\mathcal{D}_\nu\boldsymbol{\phi}\right)-\left(\mathcal{D}_\nu\boldsymbol{\phi}\right)^\dagger\sigma^a\boldsymbol{\phi}\right],\label{eqn:EoM_SU2}\\
		\partial^\mu f_{\mu\nu}&=g'j_\nu=\frac{\mathrm{i}g'}{2}\left[\boldsymbol{\phi}^\dagger\left(\mathcal{D}_\nu\boldsymbol{\phi}\right)-\left(\mathcal{D}_\nu\boldsymbol{\phi}\right)^\dagger\boldsymbol{\phi}\right],\label{eqn:EoM_U1}\\
		\mathcal{D}^\mu\mathcal{D}_\mu\boldsymbol{\phi}&=\lambda\left(\boldsymbol{\phi}^\dagger\boldsymbol{\phi}-v^2/2\right)\boldsymbol{\phi},\label{eqn:EoM_Higgs}
	\end{align}
	where $D^\mu F^a_{\mu\nu}\equiv\partial^\mu F^a_{\mu\nu}+g\varepsilon^{abc}A^{b\mu}F^c_{\mu\nu}$, and $J^a_\nu$, $j_\nu$ represent the weak isospin and hypercharge current densities~\cite{dualcurrents}, respectively.
	
	\subsection{Ansatz and its connection to baryogenesis}
	To construct the electrically neutral electroweak vortex ring configuration, we employ a generalized axially symmetric ansatz characterized by the $\phi$-winding number $n=2,\,3$. When $n=1$, the configuration reduces to the heavy string-like structure originally proposed by Nambu~\cite{Nambu}, for which high-precision numerical solutions remain elusive.
	
	We work in flat spacetime with the mostly positive metric $(-,+,+,+)$. The ansatz is expressed in spherical coordinates $(r,\theta,\phi)$, where $\theta$ is the polar angle and $\phi$ the azimuthal angle. Any expression involving Kronecker deltas is understood to be in the Cartesian basis $(x,y,z)$. The ansatz reads as follows~\cite{Teh}:
	\begin{align}
		gA^a_i=&-\frac{1}{r}\psi_1\left(r,\theta\right)\hat{n}^a_\phi\hat{\theta}_i+\frac{n}{r}\psi_2\left(r,\theta\right)\hat{n}^a_\theta\hat{\phi}_i\nonumber\\
		&+\frac{1}{r}R_1\left(r,\theta\right)\hat{n}^a_\phi\hat{r}_i-\frac{n}{r}R_2\left(r,\theta\right)\hat{n}^a_r\hat{\phi}_i,\nonumber\\
		g'B_i=&\ \frac{n}{r}B_1\left(r,\theta\right)\hat{\phi}_i,\quad gA^a_0=g'B_0=0,\nonumber\\
		\Phi^a=&\ \Phi_1\left(r,\theta\right)\hat{n}^a_r+\Phi_2\left(r,\theta\right)\hat{n}^a_\theta=H\left(r,\theta\right)\hat{\Phi}^a.\label{eqn:Ansatz}
	\end{align}
	The Higgs modulus $H\left(r,\theta\right)=\sqrt{\Phi_1^2+\Phi_2^2}$ and complex scalar doublet $\boldsymbol\phi$ are related as $\boldsymbol{\phi}=H\boldsymbol{\xi}/\sqrt{2}$, where $\boldsymbol\xi$ is a spinor that defines the Higgs field orientation in the SU(2) internal space. Specifically,
	\begin{align}
		\hat{\Phi}^a=&-\boldsymbol{\xi}^\dagger\sigma^a\boldsymbol{\xi}\nonumber\\
		=&\cos\left(\alpha-\theta\right)\hat{n}^a_r+\sin\left(\alpha-\theta\right)\hat{n}^a_\theta=h_1\hat{n}^a_r+h_2\hat{n}^a_\theta,\nonumber\\
		\boldsymbol{\xi}=&\;\mathrm{i}
		\begin{pmatrix}
			\sin\frac{\alpha\left(r,\theta\right)}{2}e^{-\mathrm{i}n\phi}\\
			-\cos\frac{\alpha\left(r,\theta\right)}{2}
		\end{pmatrix}.
	\end{align}
	The Higgs orientation is parameterized by the angle $\alpha\left(r,\theta\right)$, which approaches $2\theta$ asymptotically~\cite{Teh}. Note that $\alpha$ is not an independent parameter, as it can be expressed in terms of $\Phi_1$ and $\Phi_2$ via $\cos\alpha=(\Phi_1\cos\theta-\Phi_2\sin\theta)/H$.
	
	In the ansatz, the spatial spherical coordinate unit vectors are defined as
	\begin{align}
		\hat{r}_i&=\sin\theta\cos\phi\,\delta_{i1}+\sin\theta\sin\phi\,\delta_{i2}+\cos\theta\,\delta_{i3},\nonumber\\
		\hat{\theta}_i&=\cos\theta\cos\phi\,\delta_{i1}+\cos\theta\sin\phi\,\delta_{i2}-\sin\theta\,\delta_{i3},\nonumber\\
		\hat{\phi}_i&=-\sin\phi\,\delta_{i1}+\cos\phi\,\delta_{i2},
	\end{align}
	while the corresponding unit vectors in the isospin space are given by
	\begin{align}
		\hat{n}^a_r&=\sin\theta\cos n\phi\,\delta^a_1+\sin\theta\sin n\phi\,\delta^a_2+\cos\theta\,\delta^a_3,\nonumber\\
		\hat{n}^a_\theta&=\cos\theta\cos n\phi\,\delta^a_1+\cos\theta\sin n\phi\,\delta^a_2-\sin\theta\,\delta^a_3,\nonumber\\
		\hat{n}^a_\phi&=-\sin n\phi\,\delta^a_1+\cos n\phi\,\delta^a_2.
	\end{align}
	Here, $n$ characterizes the azimuthal winding of the field around the $z$-axis. The baryon number $Q_B$ of the configuration is ultimately determined by $n$. Driven by the chiral anomalies, the rate of change of $Q_B$ can be written as \cite{wsvortex2}
	\begin{align}
		\frac{dQ_B}{dt}&=\int d^3r \, \partial_0 j^0_B \nonumber\\
		&= \int d^3r\left\{-\partial_i j^i_B+\frac{1}{32\pi^2}\varepsilon^{\mu\nu\kappa\zeta}\left[g^2\mathrm{Tr}\left(F_{\mu\nu}F_{\kappa\zeta}\right)+\frac{1}{2}g'^2f_{\mu\nu}f_{\kappa\zeta}\right]\right\}.
	\end{align}
	In the above expression, $j^\mu_B$ is the baryon current and $F_{\mu\nu}=\sigma^aF^a_{\mu\nu}/2$. The divergence term vanishes according to Gauss's theorem, yielding
	\begin{equation}
		\frac{dQ_B}{dt} = \int d^3r\left\{\frac{1}{32\pi^2}\varepsilon^{\mu\nu\kappa\zeta}\left[g^2\mathrm{Tr}\left(F_{\mu\nu}F_{\kappa\zeta}\right)+\frac{1}{2}g'^2f_{\mu\nu}f_{\kappa\zeta}\right]\right\}=\int d^3r \, \partial_0K^0,
	\end{equation}
	where
	\begin{equation}
		K^\mu = \frac{1}{16\pi^2}\varepsilon^{\mu\nu\kappa\zeta}\left[g^2\mathrm{Tr}\left(F_{\nu\kappa}A_\zeta-\frac23\mathrm{i}gA_\nu A_\kappa A_\zeta\right)+\frac12g'^2f_{\nu\kappa}B_\zeta\right],
	\end{equation}
	is the Chern-Simons current. At a given time $t=t_0$, assuming the vacuum has $Q_B=0$ at $t=-\infty$, one can calculate the baryon number according to \cite{KlinkhamerManton}
	\begin{equation}
		Q_B = \int^{t_0}_{-\infty}dt\int d^3r \, \partial_0K^0 = \int^{t_0}_{-\infty}dt\oint_SK^idS_i+\int_{t=t_0}d^3r\,K^0.
	\end{equation}
	A particular gauge can be chosen so that $K^i$ vanishes at spatial infinity, eliminating the surface integration \cite{wsvortex1,wsvortex2}, giving
	\begin{equation}
		Q_B=\int_{t=t_0}d^3r\,K^0=\frac{n}{2}.
	\end{equation}
	
	Finally, the field equations for this configuration are obtained by substituting the ansatz into eqs.~\eqref{eqn:EoM_SU2}--\eqref{eqn:EoM_Higgs}, yielding a system of seven coupled, nonlinear, second-order partial differential equations. The explicit forms are lengthy and omitted here for brevity. The system is then solved numerically as described in the following section.

	\section{Numerical procedures}
	\subsection{Detailed computational scheme}
	We begin by introducing a dimensionless radial coordinate $x = m_W r$, where $m_W = gv/2$ is the $W$-boson mass. To further simplify the resulting system of equations, the following replacements are made:
	\begin{equation}
		H = v\widetilde{H}, \quad \lambda = g^2\beta^2, \quad g' = g\tan\theta_W.\label{eqn:dless}
	\end{equation}
	In the above expressions, the modulus and self-coupling of the Higgs field are rescaled to $\widetilde{H}$ and $\beta$, respectively, while $\theta_W$ denotes the weak mixing angle. The model parameters $\beta=0.7782$ and $\theta_W=28.18^\circ$ correspond to experimentally verified Higgs and $Z$-boson masses $m_H=125.1$ GeV and $m_Z=91.1876$ GeV~\cite{PDG}.
	
	Subsequently, the radial coordinate is compactified via $x_c = x/\left(x+1\right)$, mapping the semi-infinite domain $[0,\infty)$ to the finite interval $[0,1]$. The system of equations is then discretized on a non-uniform grid of size $160 \times 60$ spanning $0 \leq x_c \leq 1$ and $0 \leq \theta \leq \pi/2$. This exploits the reflection symmetry of the configuration about the $xy$-plane, corresponding effectively to a $160\times120$ grid over the entire domain. The truncation errors associated with the finite difference method employed are of the order $\mathcal{O}\left(\Delta x_c^2\right) \sim \mathcal{O}\left(1/160^2\right)$ radially, and $\mathcal{O}\left(\Delta\theta^2\right) \sim \mathcal{O}\left(\pi^2/120^2\right)$ in the polar direction.
	
	To capture the intricate vortex ring internal structure, the large grid size detailed above is necessary because this configuration simultaneously contains $W$- and $Z$-bosons together with the Higgs. Therefore, the transition from the unbroken ($H=0$) to the broken ($H=v$) phase is steeper, thereby causing numerical convergence difficulties. In the simpler SU(2) YMH vortex rings, only the $W$-like boson is massive. The $Z$-like boson, corresponding to the unbroken U(1), remains massless. Thus, the transition is smoother and convergence is easier to attain. As shown later, in the electroweak vortex rings, the Higgs field converges sharply within 20 radial partitions, producing a small jump in the first-order derivatives of $\Phi_1$ and $\Phi_2$, although the profile functions themselves remain smooth. An even finer grid would alleviate this, but such refinement is computationally demanding. This choice therefore represents a practical balance between numerical accuracy and available resources, and explains why high-precision electroweak vortex ring solutions have been elusive.
	
	\subsection{Boundary conditions}
	With the numerical grid established, the remaining task is to impose the boundary conditions. Asymptotically $\left(r\rightarrow\infty\right)$,
	\begin{align}
		\psi_A\left(\infty,\theta\right)&=2,\ R_A\left(\infty,\theta\right)=B_1\left(\infty,\theta\right)=0,\nonumber\\
		\Phi_1\left(\infty,\theta\right)&=\cos\theta,\ \Phi_2\left(\infty,\theta\right)=\sin\theta,\label{eqn:BCrinf}
	\end{align}
	where $A=1,2$. At the origin $\left(r=0\right)$, 
	\begin{align}
		&\psi_A\left(0,\theta\right)=R_A\left(0,\theta\right)=B_1\left(0,\theta\right)=0,\nonumber\\
		&\Phi_1\left(0,\theta\right)\sin\theta+\Phi_2\left(0,\theta\right)\cos\theta=0,\nonumber\\
		&\partial_r\left(\Phi_1\left(r,\theta\right)\cos\theta-\Phi_2\left(r,\theta\right)\sin\theta\right)\big|_{r=0}=0.\label{eqn:BCr0}
	\end{align}
	On the positive $z$-axis $\left(\theta=0\right)$,
	\begin{align}
		\partial_\theta\psi_A=R_A=\partial_\theta\Phi_1=\Phi_2=B_1=0,\label{eqn:BCtheta0}
	\end{align}
	and on the equatorial plane $\left(\theta=\pi/2\right)$,
	\begin{align}
		\partial_\theta\psi_A=R_A=\Phi_1=\partial_\theta\Phi_2=\partial_\theta B_1=0.\label{eqn:BCequator}
	\end{align}
	
	The boundary conditions specified above are constructed based on the seminal work of Kleihaus and Kunz \cite{KKMAP} and later contributions \cite{ymhvortex,Amin}. Specifically, those for the six profile functions associated with the SU(2) gauge fields and the Higgs field match the setup in Ref. \cite{Amin} exactly. For the remaining U(1) gauge field profile function, $B_1(r,\theta)$, its boundary conditions are derived analytically by substituting the asymptotic behaviors of the other six profile functions into the U(1) field equation at the boundaries. The same boundary conditions for the electroweak vortex ring configurations have also been successfully utilized in recent literature \cite{Teh,ZhuDan1}.

	\subsection{Numerical method and initial guesses}
	The electroweak vortex ring solutions are then obtained numerically using MATLAB, with the boundary conditions fixed as detailed in the previous section. We employ the \texttt{fsolve} package utilizing a trust-region-reflective algorithm, supplied with a Jacobian sparsity pattern. For our setup, the Jacobian, $J(\mathbf{x})$, is a $65667\times65667$ matrix---corresponding to the seven profile functions and equations, each partitioned into $159\times59$ internal data points ($7\times159\times59=65667$). This binary sparsity pattern significantly enhances computational efficiency, and convergence is achieved in just a few hundred iterations if an appropriate initial guess is provided. The convergence criteria are defined by the residual norm and the first-order optimality measure, which will be discussed in the following section.
	
	For the numerical solutions to converge, suitable initial guesses are imperative. In this work, we first reproduce the well-known SU(2) YMH vortex ring solutions \cite{ymhvortex,Amin}. These are then converted into initial guesses for the electroweak system by appending a segment of zeros to represent the profile function $B_1(r,\theta)$ in the ansatz. Nevertheless, the electroweak vortex ring configuration proves to be extremely sensitive to initial conditions. Through extensive numerical experimentation, we find that convergence is only achieved when the underlying SU(2) vortex ring solution itself possesses a sufficiently large Higgs self-coupling $\beta$. This requirement clarifies why the seemingly simpler $n=1$ case remains numerically inaccessible: the heavy string-like structure proposed by Nambu~\cite{Nambu} lacks a direct counterpart in the SU(2) YMH theory, and consequently, a suitable initial guess cannot be constructed.
	
	\subsection{Convergence tests}
	Our numerical solutions exhibit excellent precision, as quantified by the residual norm and the first-order optimality measure. For the discretized system of equations, \texttt{fsolve} finds a solution to $\mathbf{F}(\mathbf{x})=\mathbf{0}$, where $\mathbf{F}$ is the residual vector of the system in the current iteration and $\mathbf{x}$ is a column vector of 65667 elements (flattened from the seven two-dimensional profile functions). The squared residual norm, $f(\mathbf{x})^2$, is defined as the sum of squares of the values of $\mathbf{F}$, $f(\mathbf{x})^2 = \|\mathbf{F}(\mathbf{x})\|^2 = \sum_i F_i(\mathbf{x})^2$, while the first-order optimality measure, TolX, calculates the infinity norm of $J(\mathbf{x})^T \mathbf{F}(\mathbf{x})$, where $J(\mathbf{x})$ is the Jacobian. The first converged solution within the Weinberg-Salam model was obtained for parameters $n=3$, $\beta=2$, and $\theta_W = 28.18^\circ$, yielding $f(\mathbf{x})^2 = 7.2745 \times 10^{-20}$ and TolX $= 1.9760 \times 10^{-5}$. This high-precision solution served as the initial guess for generating all subsequent solutions at different parameter points.
	
	A systematic observation is that the convergence metrics naturally degrade for solutions obtained further away from the initial guess in parameter space, particularly for those with the Higgs self-coupling $\beta > 20$ (and therefore, discarded). This is an expected feature of the numerical process. Nevertheless, the quality of all solutions remains excellent. Specifically:
	\begin{itemize}
		\item For the physical $n=2$ vortex ring ($\beta = 0.7782$, $\theta_W = 28.18^\circ$): $f(\mathbf{x})^2 = 8.9387 \times 10^{-14}$, TolX $= 9.6729 \times 10^{-5}$.
		\item For the physical $n=3$ vortex ring ($\beta = 0.7782$, $\theta_W = 28.18^\circ$): $f(\mathbf{x})^2 = 7.6142 \times 10^{-14}$, TolX $= 3.3186 \times 10^{-4}$.
	\end{itemize}
	Despite the increase in these residual metrics, all field profiles and derived physical quantities (e.g., total energy, current density distributions, etc.) are smooth and well-resolved, confirming the robustness of the solutions.

	\section{Vortex ring properties}
	\subsection{EM and neutral field lines}
	In the Weinberg-Salam model, the physical electromagnetic and $Z$-boson fields are related to the fundamental SU(2) and U(1) gauge fields via the standard rotation by the Weinberg angle $\theta_W$:
	\begin{align}
		A_i^{\text{em}} &= \sin\theta_WA_i^{'3}+\cos\theta_WB_i,\\
		Z_i &= \cos\theta_WA_i^{'3}-\sin\theta_WB_i.
	\end{align}
	A form more directly aligned with our ansatz, eq.~\eqref{eqn:Ansatz}, is obtained by multiplying with the couplings $g$, $g'$, and the elementary electric charge $e=gg'/\sqrt{g^2+g'^2}$:
	\begin{align}
		eA_i^{\text{em}}&=\sin^2\theta_W\left(gA_i^{'3}\right)+\cos^2\theta_W\left(g'B_i\right),\label{eqn:AiMix}\\
		eZ_i&=\cos\theta_W\sin\theta_W\left[\left(gA_i^{'3}\right)-\left(g'B_i\right)\right].\label{eqn:ZiMix}
	\end{align}
	
	The specific form of the SU(2) gauge potential $gA^{'3}_i$ entering these expressions is obtained by applying the unitary gauge that rotates the Higgs doublet to $(0\ v)^\mathrm{T}/\sqrt{2}$ \cite{Teh,ZhuDan3}. Within our model, it reads:
	\begin{equation}
		gA^{'3}_i=\frac{n}{r}\left(\psi_2h_2-R_2h_1-\frac{1-\cos\alpha}{\sin\theta}\right)\hat{\phi}_i=\frac{n}{r}A_1\hat{\phi}_i.
	\end{equation}
	The terms in parentheses are collectively denoted as $A_1$, introduced for notational convenience. The ``magnetic" part of the gauge fields can then be calculated as
	\begin{align}
		g'\mathcal{B}^{\scalebox{.5}{\mbox{U(1)}}}_i&=-\frac{1}{2}\varepsilon_{ijk}\left(g'f_{jk}\right)=-n\varepsilon_{ijk}\partial_j\left\{B_1\sin\theta\right\}\partial_k\phi,\\
		g\mathcal{B}^{\scalebox{.5}{\mbox{SU(2)}}}_i&=-\frac{1}{2}\varepsilon_{ijk}\left(gF_{jk}\right)=\varepsilon_{ijk}\partial_j\left(gA'^3_k\right)=-n\varepsilon_{ijk}\partial_j\left\{A_1\sin\theta\right\}\partial_k\phi,
	\end{align}
	where $F_{jk}=\partial_jA'^3_k-\partial_kA'^3_j$ is identical to the field strength originally proposed by 't Hooft \cite{Teh,tHooft}. Following the mixings shown in eqs.~\eqref{eqn:AiMix}\&\eqref{eqn:ZiMix}, the physical magnetic field and the ``magnetic" part of the $Z$-boson field are
	\begin{align}
		e\mathcal{B}^\text{em}_i&=-\frac{1}{2}\varepsilon_{ijk}\left[\sin^2\theta_W\left(gF_{jk}\right)+\cos^2\theta_W\left(g'f_{jk}\right)\right]\label{eqn:em_field_lines}\nonumber\\
		&=-n\varepsilon_{ijk}\partial_j\left\{\sin^2\theta_WA_1\sin\theta+\cos^2\theta_WB_1\sin\theta\right\}\partial_k\phi,\\
		e\mathcal{B}^Z_i&=-\frac{1}{2}\cos\theta_W\sin\theta_W\varepsilon_{ijk}\left[\left(gF_{jk}\right)-\left(g'f_{jk}\right)\right]\nonumber\\
		&=-n\cos\theta_W\sin\theta_W\varepsilon_{ijk}\partial_j\left\{\left(A_1-B_1\right)\sin\theta\right\}\partial_k\phi.\label{eqn:Z_field_lines}
	\end{align}
	Their respective field lines are constructed by drawing the contour lines of the terms enclosed in curly brackets, for $\phi=0$, i.e. in the meridian plane.
	
	\subsection{EM, $Z$-, and $W$-boson currents}\label{subsec:currents}
	To provide a comprehensive visualization of the internal structures, we define the electromagnetic and weak neutral currents in terms of the charge operators $Q_\text{em} = T_3 + Y_W/2$ and $Q_Z = T_3 - Q_\text{em} \sin^2\theta_W$~\cite{chargeoperator},
	\begin{align}
		j_i^{\text{em}} &= J_i^3+\frac{1}{2}j_i,\label{eqn:em_current}\\
		j_i^{Z} &= J_i^3\cos^2\theta_W-\frac{1}{2}j_i\sin^2\theta_W,\label{eqn:z_current}
	\end{align}
	where $J_i^a$, $j_i$ are the weak isospin and hypercharge current densities~\cite{dualcurrents}, respectively. In addition, the charged currents for the $W$-bosons are
	\begin{equation}
		j_i^{W^\pm}=\frac{J_i^1\mp\mathrm{i}J_i^2}{\sqrt{2}}.\label{eqn:wcurrent}
	\end{equation}
	
	The EM and $Z$-boson currents, obtained through direct substitution of the ansatz into eqs.~\eqref{eqn:em_current}\&\eqref{eqn:z_current}, are purely azimuthal. In contrast, the $W$-boson currents exhibit a far richer spatial structure. Their analytical form, derived from eq.~\eqref{eqn:wcurrent},
	\begin{align}
		j_i^{W^\pm}=-&\frac{H^2}{4\sqrt{2}r}e^{\mp\mathrm{i}n\phi}\bigg\{n\big[R_2\sin\theta-\psi_2\cos\theta+\left(B_1+\csc\theta\right)\sin\alpha\big]\hat\phi_i\nonumber\\
		&\hspace{5.4em}\pm e^{\mathrm{i}\pi/2}\left[\left(R_1+\alpha'r\right)\hat{r}_i+\left(\dot\alpha-\psi_1\right)\hat\theta_i\right]\bigg\},\label{eqn:wcurrentanalytical}
	\end{align}
	is inherently complex, reflecting the charged nature of $W$-bosons. Note that in the above expression, the $(\hat{r}_i,\hat\theta_i)$ and $\hat\phi_i$ components are in quadrature, giving rise to a nontrivial spatial flow pattern with an azimuthally varying direction. This structure is reminiscent of the knot-like configurations studied in classical field theory~\cite{Hopfion1,Hopfion2}, which is natural given that the underlying Higgs field realizes the first Hopf map, $\Phi^a:S^3\mapsto S^2$.
	
	\subsection{Energy density and total energy}\label{subsec:energy}
	For the electroweak vortex ring configuration studied here, the energy density is calculated as $\varepsilon=T_{00}=g_{00}\mathcal{L}=-\mathcal{L}$, see eq.~\eqref{eqn:T}. Upon performing the dimensionless transformation, eq.~\eqref{eqn:dless}, the energy density becomes
	\begin{align}
		\varepsilon=&\,\frac{1}{4g^2}\left(gF^a_{ij}\right)\left(gF^{aij}\right)+\frac{1}{4g'^2}\left(g'f_{ij}\right)\left(g'f^{ij}\right)+\left(\mathcal{D}_i\boldsymbol{\phi}\right)^\dagger\left(\mathcal{D}^i\boldsymbol{\phi}\right)+\frac{\lambda}{2}\left(\boldsymbol{\phi}^\dagger\boldsymbol{\phi}-v^2/2\right)^2\nonumber\\
		=&\,\frac{m_W^2v^2}{4}\widetilde{F}^a_{ij}\widetilde{F}^{aij}+\frac{m_W^2v^2}{4\tan^2\theta_W}\widetilde{f}_{ij}\widetilde{f}^{ij}\nonumber\\
		&+m_W^2v^2\left(\widetilde{\mathcal{D}}_i\widetilde{H}\boldsymbol{\xi}\right)^\dagger\left(\widetilde{\mathcal{D}}^i\widetilde{H}\boldsymbol{\xi}\right)+m_W^2v^2\frac{\beta^2}{2}\left(\widetilde{H}^2-1\right)^2=m_W^2v^2\widetilde{\varepsilon}.\label{eqn:energy_density}
	\end{align}
	The dimensionless energy density $\widetilde{\varepsilon}$ is defined by factoring out the scale $m_W^2v^2$. Notably, the $\widetilde{f}_{ij}\widetilde{f}^{ij}$ term introduces a factor of $1/\tan^2\theta_W$. When accounting for the $\cos\theta_W\sin\theta_W$ factor in eq.~\eqref{eqn:ZiMix}, the energy density contribution from the U(1) component of the $Z$-boson field is modulated by $\cos\theta_W\sin\theta_W/\tan^2\theta_W$.
	
	Integration of eq.~\eqref{eqn:energy_density} over all space yields the mass of the electroweak vortex ring in units of TeV:
	\begin{align}
		E &= \iint r^2\sin\theta\;\varepsilon\;dr\;d\theta\int d\phi = 2\pi\iint \left(\frac{x^2}{m_W^2}\right)\sin\theta\left(m_W^2v^2\,\widetilde\varepsilon\right)\frac{dx}{m_W}d\theta\nonumber\\
		&=\left(\frac{2\pi v^2}{m_W}\right)\iint x^2\sin\theta\;\widetilde\varepsilon\; dx\;d\theta=\left(\frac{2\pi v^2}{m_W}\right)\widetilde{E}.\label{eqn:efactor}
	\end{align}
	By employing $m_W=80.379\ \text{GeV}$~\cite{PDG} and $v=(\sqrt{2}G_F)^{-1/2}\approx246.2174\ \text{GeV}$~\cite{HiggsVEV}, with the Fermi constant $G_F=1.1664\times10^{-5}\ \text{GeV}^{-2}$~\cite{Fermi}, the prefactor, $2\pi v^2/m_W$, is approximately 4.7389 TeV.
	
	Other physical quantities investigated, such as the stress-energy tensor components $T_{33}$ (for $T_{11}$, $\phi$ is set to 0) studied in the following section, are obtained by direct substitution of the ansatz into their respective definitions. For the sake of brevity, their explicit analytical forms are omitted.

	\section{Results and discussion}
	\subsection{Higgs modulus}\label{sec:HiggsModulus}
\begin{figure}[t]
		\includegraphics[width=\textwidth]{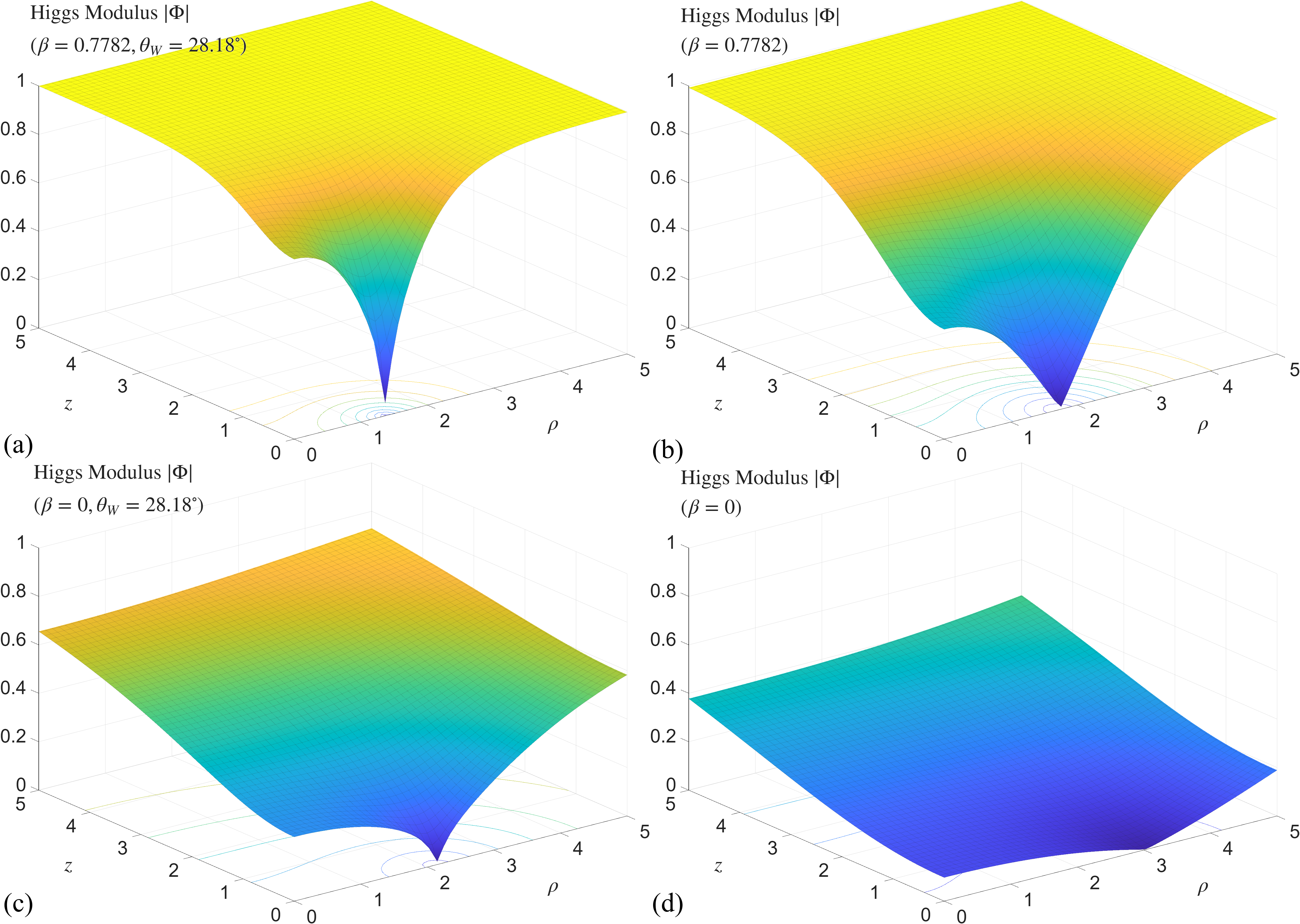}
		\caption{3D Higgs modulus surface plots for (a, c) electroweak and (b, d) SU(2) YMH vortex ring solutions at different Higgs self-coupling constant $\beta$: (a, b) physical Higgs mass and (c, d) BPS limit. In the case of the electroweak vortex rings, the Weinberg angle is always physical $\theta_W=28.18^\circ$.}
		\label{fig:Higgs_Modulus}
	\end{figure}

	Figure~\ref{fig:Higgs_Modulus} presents the three-dimensional visualizations of the Higgs modulus for selected solutions using cylindrical coordinates.
	The parameter $\rho$ denotes radial distance and is defined as $\rho \equiv \sqrt{x^2+y^2}$. In the vicinity of the vortex ring, where $H=|\Phi|=0$, the Higgs modulus of the electroweak vortex rings approaches zero more rapidly than that of their SU(2) counterparts. This is evident from the narrower and more pointed inverted cones of panels (a,c) relative to (b,d). In the case of the SU(2) vortex ring when the Higgs self-coupling $\beta=0$, figure~\ref{fig:Higgs_Modulus}(d), the cross-section when $z=0$ is essentially flat, whereas the corresponding electroweak version in panel (c) exhibits a clearly identifiable pointed dip. According to boundary condition~\eqref{eqn:BCequator}, $\Phi_1=0$ in the equatorial plane. Consequently, the behavior of Higgs modulus in the $xy$-plane is solely determined by $\Phi_2$, and the pointed dip leads to the aforementioned jump in $\partial\Phi_2/\partial r$. An analogous argument holds for $\partial\Phi_1/\partial\theta$. These moderate discontinuities can be smoothed out by increasing grid size. A rough estimate is that $r$ partition needs to exceed 500 and $\theta$ partition to 180.
	
	At spatial infinity, the Higgs modulus of electroweak vortex rings approaches the vacuum expectation value significantly faster than that of their SU(2) counterparts. In figure~\ref{fig:Higgs_Modulus}(d), along the $z$-axis at $z=5\,m_W^{-1}$, the normalized Higgs modulus for an SU(2) vortex ring barely reaches 0.4, whereas for the electroweak version in panel (c), it has already exceeded 0.6. This reflects that the additional massive $Z$-boson causes the steeper transition from unbroken ($H=0$) to broken ($H=v$) phase stated earlier, explaining the root of the numerical challenges.
	
	\subsection{Geometric size}
	The stabilization paradigm recently identified in Cho-Maison MAPs \cite{ZhuDan3}, featuring both Higgs- and $Z$-boson-mediated repulsions, extends to the electroweak vortex rings studied here. The crucial distinction is the absence of a topological monopole charge. The persistence of these repulsive interactions in the charge-free vortex ring configuration demonstrates that they are fundamental to the Weinberg-Salam model and not contingent upon topological structures. The strength and interplay of these repulsive interactions directly determine the geometric size of the vortex rings, which is quantified by the radius $R_\rho$, defined as the distance from the origin to the local minimum of Higgs modulus $\widetilde{H}$.
	
\begin{figure}[t]
		\includegraphics[width=\textwidth]{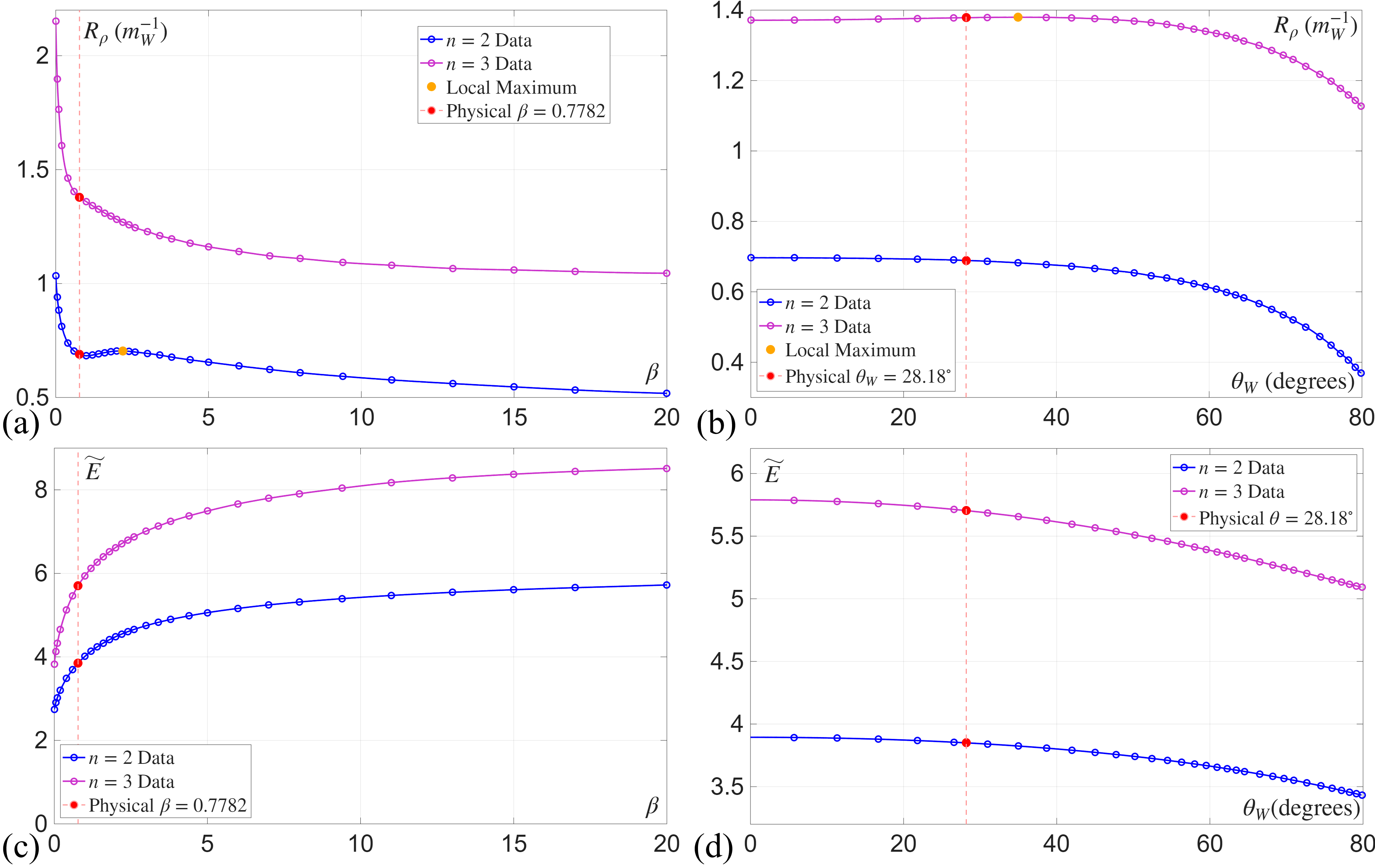}
		\caption{Dependence of electroweak vortex ring properties on model parameters.  
			(a,b) Radius $R_\rho$ (in units of $m_W^{-1}$) as a function of the Higgs self-coupling $\beta$ and Weinberg angle $\theta_W$.  
			(c,d) Dimensionless total energy $\widetilde{E}$ versus $\beta$ and $\theta_W$. Results are shown for $\phi$-winding numbers $n=2,\,3$. Physical model parameters correspond to $\beta=0.7782$ and $\theta_W=28.18^\circ$ (vertical dotted lines).}
		\label{fig:R_E_Trends}
	\end{figure}

	The dependence of $R_\rho$ on both Higgs self-coupling $\beta$ and Weinberg angle $\theta_W$ is presented in figure~\ref{fig:R_E_Trends}(a,b), where solutions with $\phi$-winding number $n = 2,\,3$ are shown.
	The observed trends are governed by the mass-controlled ranges of the repulsive interactions. Increasing $\beta$ raises the Higgs mass $m_H$, shortening the range of the repulsion and thereby, reducing $R_\rho$. Notably, the $n=2$ data produces a local maximum in figure~\ref{fig:R_E_Trends}(a), which deviates from the monotonic exponential form of a Yukawa potential, signaling the complexity and non-linearity of the Higgs-mediated repulsion. Similarly, increasing $\theta_W$ raises the $Z$-boson mass $m_Z$ (with $m_Z=m_W$ at $\theta_W=0$) and accounts for the decrease in $R_\rho$, with a subtle non-monotonic feature discernible in figure~\ref{fig:R_E_Trends}(b). Thus, in the absence of a topological charge, these repulsive interactions operate as the sole sculptors of the size of the vortex rings.
	
	\subsection{Mass predictions}	
	Extending this logic, the total energy $E$ of the configuration is likewise governed by the repulsive interactions. A key advantage of the electroweak vortex ring over the singular Cho-Maison MAP is its regularity, which ensures a finite $E$. The total energy then offers a complementary and singularity-free perspective on the stabilization paradigm. Crucially, with the Weinberg angle as a free parameter, our numerical model enables the first direct calculation of its physical mass in TeV.
	
	Figure~\ref{fig:R_E_Trends}(c,d) shows the dimensionless total energy $\widetilde{E}$ as a function of $\beta$ and $\theta_W$ for the $n = 2,\,3$ vortex rings. Conceptually, increasing the mediator mass enhances the interaction strength while reducing its range. As a result, the configuration becomes more compact, the internal interactions intensify, and the total energy increases. This behaviour is evident from figure~\ref{fig:R_E_Trends}(a,c), where $\widetilde{E}$ grows with $\beta$, while $R_\rho$ shrinks accordingly. One would expect the same behaviour for the $Z$-boson repulsion: increasing $\theta_W$ raises $m_Z$, which would lead to an increase in $\widetilde{E}$. However, figure~\ref{fig:R_E_Trends}(b,d) shows the opposite, which does not constitute a contradiction. As stated in section~\ref{subsec:energy}, the U(1) contribution to $\widetilde{E}$ from the $Z$-boson is modulated by $\cos\theta_W\sin\theta_W/\tan^2\theta_W$, which decreases monotonically as $\theta_W \to 90^\circ$, explaining the observed downward trend of $\widetilde{E}$.
	
	From eq.~\eqref{eqn:efactor}, the total energy $E$ relates to $\widetilde{E}$ via
	\begin{equation}
		E = \frac{2\pi v^2}{m_W} \widetilde{E} \approx (4.7389\ \text{TeV}) \times \widetilde{E}.
	\end{equation}
	For physical model parameters (vertical dotted lines in figure~\ref{fig:R_E_Trends}), $\widetilde{E}_{(n=2)}=3.8001$ and $\widetilde{E}_{(n=3)}=5.6560$, yielding
	\[
	E_{(n=2)} \approx 18.01\ \text{TeV}, \qquad E_{(n=3)} \approx 26.80\ \text{TeV}.
	\]
	By comparison, the lightest $n=1$ configuration is expected to have $E<14$ TeV, theoretically within reach of the Large Hadron Collider (LHC), but with minuscule cross-section. Definitive experimental signatures and the exploration of the more massive $n\geq2$ states reported here will likely require the proposed Future Circular Collider (FCC-hh).
	
	\subsection{Field lines}
\begin{figure}[t]
		\includegraphics[width=\textwidth]{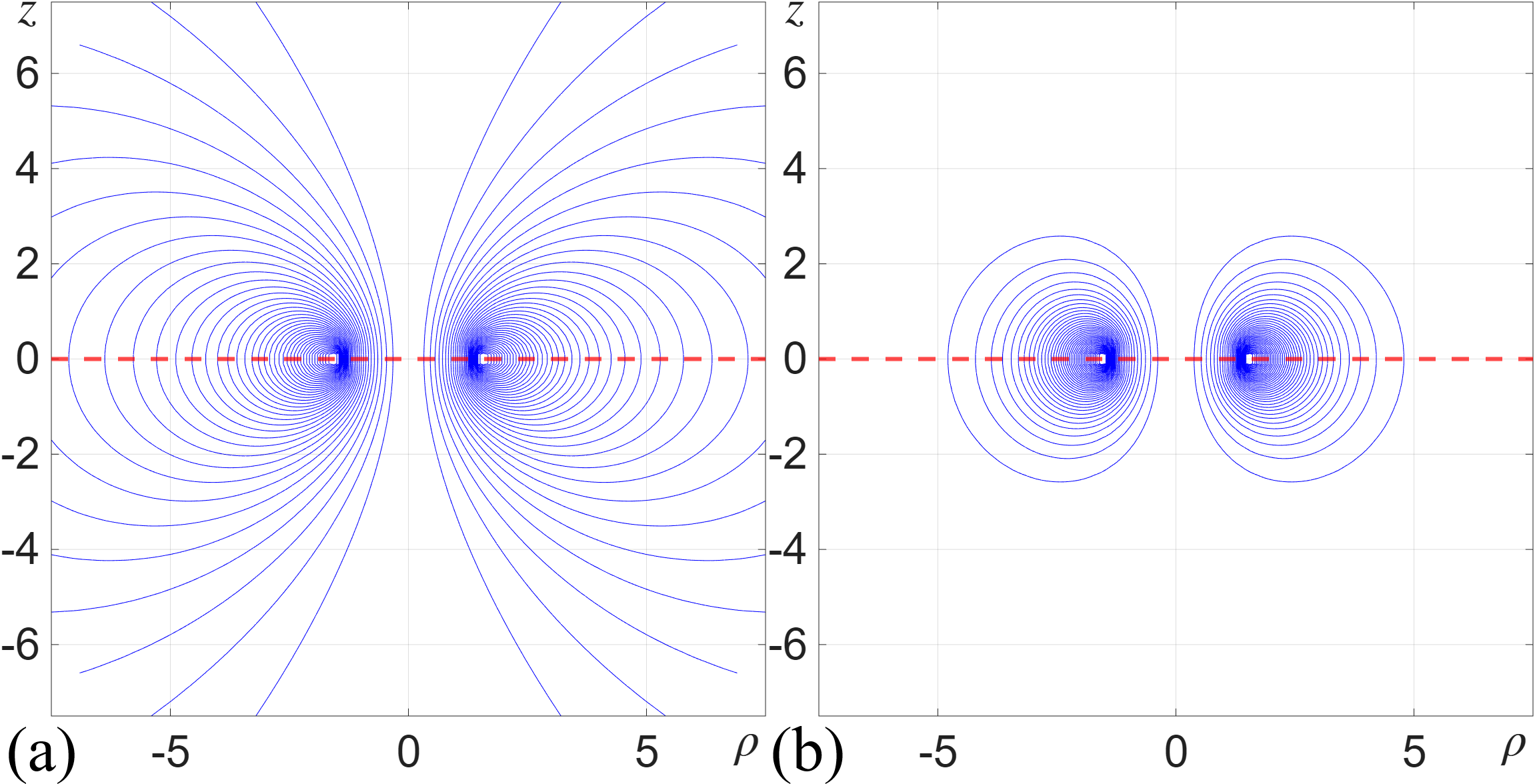}
		\caption{Field lines of (a) the electromagnetic field and (b) the neutral $Z$-boson field for the electroweak vortex ring solution with $n = 3$, $\beta = 0.7782$, and $\theta_W = 28.18^\circ$.}
		\label{fig:field_lines}
	\end{figure}

	Having established the TeV-scale masses of the vortex rings, we now examine their internal structures. Figure~\ref{fig:field_lines} depicts the electromagnetic and $Z$-boson field lines for the physical vortex ring solution with $n = 3$, $\beta = 0.7782$, and $\theta_W = 28.18^\circ$.
	Both sets of field lines form concentric circles. However, a critical distinction emerges: while the massless photon field extends to spatial infinity in panel (a), the massive $Z$-boson field lines are confined within $|\rho|<5\,m_W^{-1}$ (with $\rho$ defined in section~\ref{sec:HiggsModulus}), providing a direct visualization of its finite interaction range, as shown in figure~\ref{fig:field_lines}(b).
	
	The similar geometry of the electromagnetic and $Z$-boson field lines reflects an analogous relation between their respective field and current distributions. In the electromagnetic case, the concentric magnetic field is associated with a circulating electric current via the classical Amp\`ere's circuital law. For the $Z$-boson field, the corresponding relation follows directly from the electroweak field equation, eq.~\eqref{eqn:EoM_SU2}, which links the curl of the non-Abelian magnetic field to the Higgs current. Thus, the $Z$-boson field-line configuration is a manifestation of a neutral analogue of the familiar Amp\`ere's circuital law: a circulating azimuthal ``electric" neutral current is associated with a concentric ``magnetic" neutral field in the meridian plane.
	
	\subsection{Current distributions}
\begin{figure}[t]
		\includegraphics[width=\textwidth]{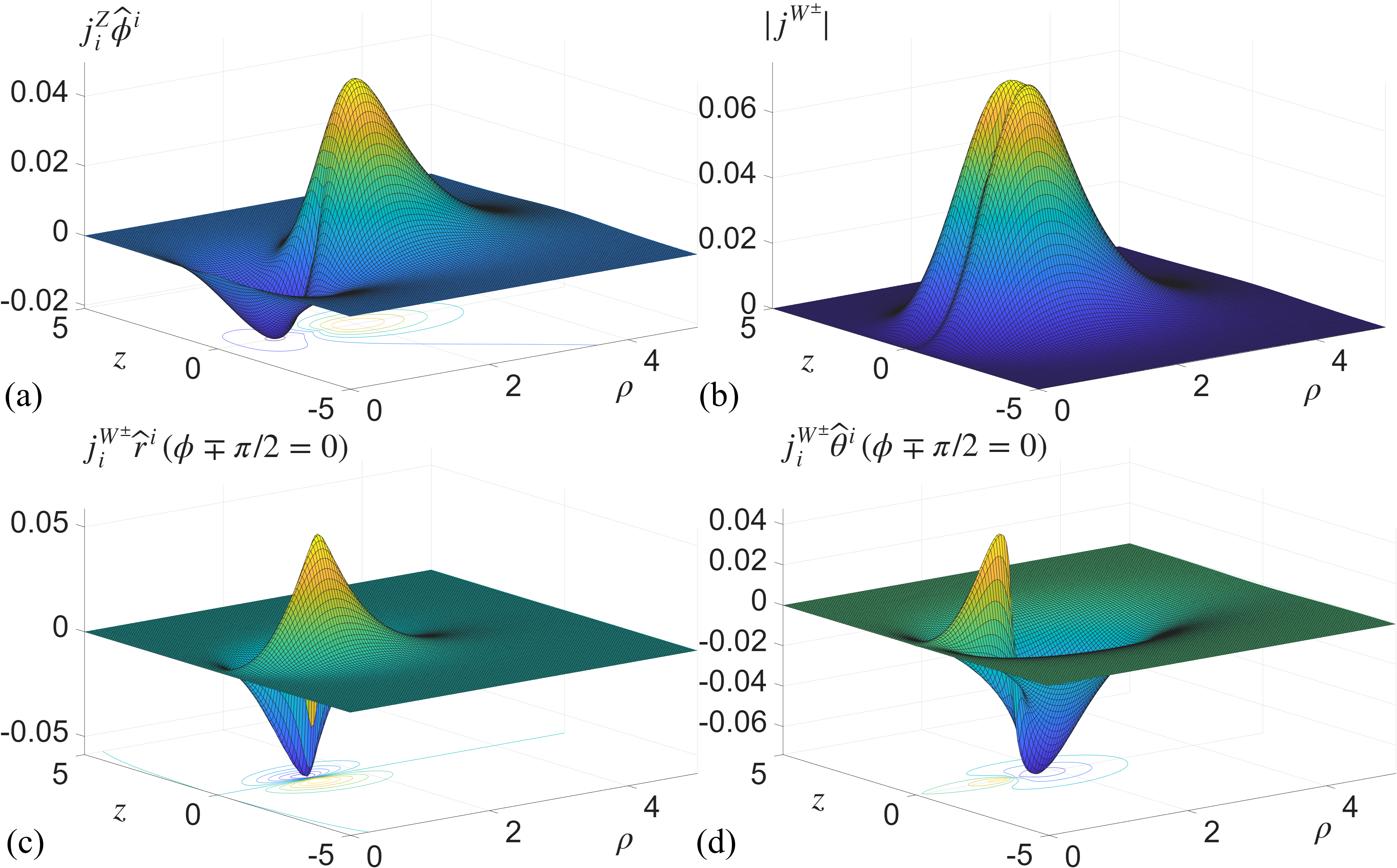}
		\caption{Spatial distributions of (a) azimuthal neutral current $j_i^Z\hat\phi^i$, (b) modulus $|j^{W^\pm}|$, (c) radial component $j^{W^\pm}_i\hat{r}^i$, and (d) polar component $j^{W^\pm}_i\hat{\theta}^i$ of the $W$-boson currents for the physical electroweak vortex ring (same parameters as in figure~\ref{fig:field_lines}).}
		\label{fig:currents}
	\end{figure}

	In accordance with the concentric $Z$-boson field lines discussed above, the neutral current derived analytically from eq.~\eqref{eqn:z_current} is purely azimuthal. Figure~\ref{fig:currents}(a) shows the spatial distribution of $j_i^Z\hat\phi^i$, revealing both positive and negative regions.
	This indicates two concentric current loops circulating in opposite directions, with the outer ring carrying a higher current density. Though not anticipated, this double-loop structure is consistent with the relation between the $Z$-boson field and the Higgs current dictated by the electroweak field equations. The electromagnetic current shows a similar distribution.
	
	As stated in section~\ref{subsec:currents}, the $W$-boson currents are inherently complex, making their spatial structure less straightforward to visualize. We therefore examine their distribution through the quantities presented in figure~\ref{fig:currents}(b)--(d). The modulus $|j^{W^\pm}|$ shown in panel~(b) reveals a clear double-torus structure, with maxima located symmetrically about the $xy$-plane at $z\approx\pm0.34$. Thus, the $W$-boson currents are concentrated around two parallel tori, in stark contrast to the concentric $j_i^\text{em}$ and $j_i^Z$. A numerical decomposition of $j_i^{W^\pm}$ reveals that the azimuthal component has opposite signs in the upper and lower hemisphere and does not form a conventional circulating current. Rather, the flow alternates in direction around the toroidal structure, producing $2n$ azimuthal lobes, like segments of a peeled tangerine. This behavior originates from the overall phase factor $e^{\mp\mathrm{i}n\phi}$ in the analytical expression for $j_i^{W^\pm}$, eq.~\eqref{eqn:wcurrentanalytical}. The real part has a $\cos(n\phi)$ dependence, and consequently, the azimuthal current changes sign at $\phi=(2k+1)\pi/(2n),\,k=0,1,\ldots,2n-1$, dividing the toroidal structure into $2n$ azimuthal lobes with alternating flow directions.
	
	Figure~\ref{fig:currents}(c) and (d) complete the picture by showing the corresponding radial and polar components of the $W$-boson currents at $\phi\mp\pi/2=0$. This choice eliminates the phase factor from these components, allowing their spatial structures to be displayed directly. In panel~(c), the radial component also has opposite signs in the two hemispheres, indicating inward and outward flows above and below the $xy$-plane. In panel~(d), the polar component reverses sign as $\rho$ increases: it is positive near the origin and negative at larger $\rho$. The polar component roughly corresponds to flow along the $z$-axis, so at $\phi\mp\pi/2=0$, the inner region flows upward while the outer region sinks. Note that both the radial and polar components are modulated by the overall phase factor and are therefore periodic in $\phi$. As the azimuthal angle varies from $0$ to $2\pi$, each component changes sign $2n$ times, producing the alternating flow pattern around the toroidal structure.
	
	\subsection{Mechanical stresses}
	Having established the internal structure of the electroweak vortex rings, we now examine the stress distributions within this configuration. In our previous study of Cho-Maison MAPs~\cite{ZhuDan3}, the stress-energy tensor analysis served as a key diagnostic for identifying the repulsive interactions. Here, we adopt the same approach to probe the horizontal and vertical forces acting on the vortex ring via $T_{11}$ and $T_{33}$. Unlike the Cho-Maison MAP, the electroweak vortex ring is regular, and integrating these components reveals the net stress experienced by the configuration.
	
\begin{figure}
		\includegraphics[width=\columnwidth]{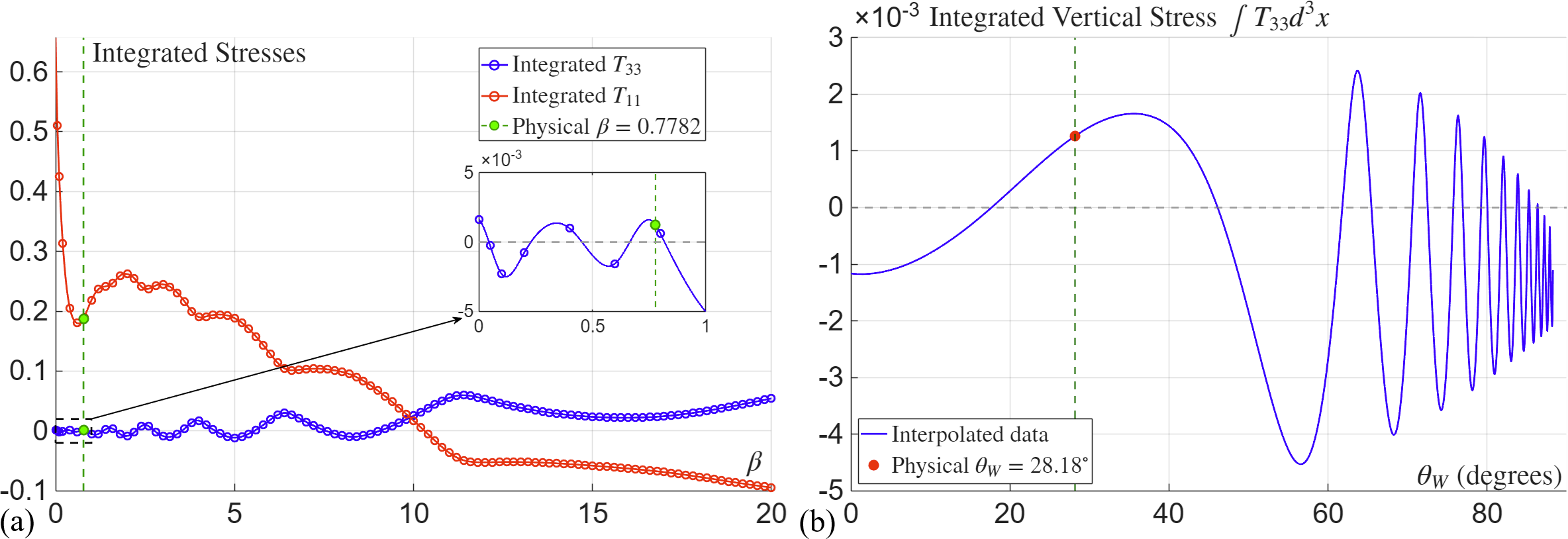}
		\caption{The integrated stresses for $n = 3$ electroweak vortex ring solutions versus $\beta$ and $\theta_W$. Vertical dotted lines correspond to the physical $\beta=0.7782$ and $\theta_W=28.18^\circ$. In (a) both horizontal ($T_{11}$) and vertical ($T_{33}$) stresses are shown against $\beta$, whereas in (b) only the vertical stress is plotted against $\theta_W$.}
		\label{fig:t11t33}
	\end{figure}

	Figure~\ref{fig:t11t33} shows the integrated stresses of the $n=3$ electroweak vortex rings as a function of $\beta$ and $\theta_W$, based on over 200 high-precision data points.
	In figure~\ref{fig:t11t33}(a), both curves display pronounced oscillatory behaviour. Notably, the integrated $T_{33}$ alternates between repulsive and attractive regimes as the Higgs mass increases, changing signs four times in the interval $0\leq\beta\leq0.7782$ (see inset of figure~\ref{fig:t11t33}(a)), corresponding to the range from the BPS limit to the physical Higgs mass. For the $\theta_W$ dependence shown in panel (b), only the integrated $T_{33}$ is shown because the horizontal stress lacks a significant oscillation and the scale difference is large (integrated $T_{33}$ values are of order $10^{-3}$). Including the horizontal component would render the oscillatory features imperceptible. Furthermore, only the interpolated curve is displayed; the discrete data points are omitted to avoid obscuring the behaviour of the curve. Overall, the oscillatory nature of these stresses points to an inherent instability, and it is plausible that the variation of the internal forces could influence the decay channels of these topological structures, potentially leaving a distinctive imprint on the electroweak epoch.
	
	Intuitively, it is surprising that a configuration governed by repulsive interactions exhibits an attractive phase. However, this behavior can be understood by considering the internal current structures. In plasma physics, a conducting filament generates a magnetic field that exerts an inward Lorentz force back onto the current distribution---a phenomenon known as the Bennett pinch effect. The oscillatory curves in figure~\ref{fig:t11t33} thus represent the competition between Higgs-, $Z$-boson-mediated repulsions and the ``pinching" pressure from previously described current distributions.
	
	\subsection{Neutral Bennett pinch}	
	While the connection between the neutral analogue of Amp\`ere's circuital law, the electroweak field equations, and the field-line configurations is evident from our numerical solutions, identifying a corresponding Bennett pinch for the neutral field remains preliminary. This is because in the electroweak sector, the stress-energy tensor components $T_{ii}$ are complex composite quantities. 
	
	Specifically, the local mechanical balance captured by $T_{11}$ and $T_{33}$ simultaneously incorporates several competing contributions:
	\begin{enumerate}
		\item \textbf{Repulsive interactions} arising from the Higgs and $Z$-boson fields, as established in the stabilization paradigm~\cite{ZhuDan3};
		\item \textbf{Electromagnetic Bennett pinch} (attractive), which is a standard consequence of the U(1) electromagnetic current loops;
		\item A hypothesized \textbf{neutral Bennett pinch} (attractive), potentially generated by the concentric ``magnetic" neutral field.
	\end{enumerate}
	Due to the intrinsic non-linear coupling of the Weinberg-Salam model, these terms cannot be cleanly isolated within our current numerical framework to provide a conclusive deduction of the neutral pinch's specific magnitude. Nevertheless, our numerical solutions show promising evidence pointing towards such an effect for the neutral field.
	
\begin{figure}
		\includegraphics[width=\textwidth]{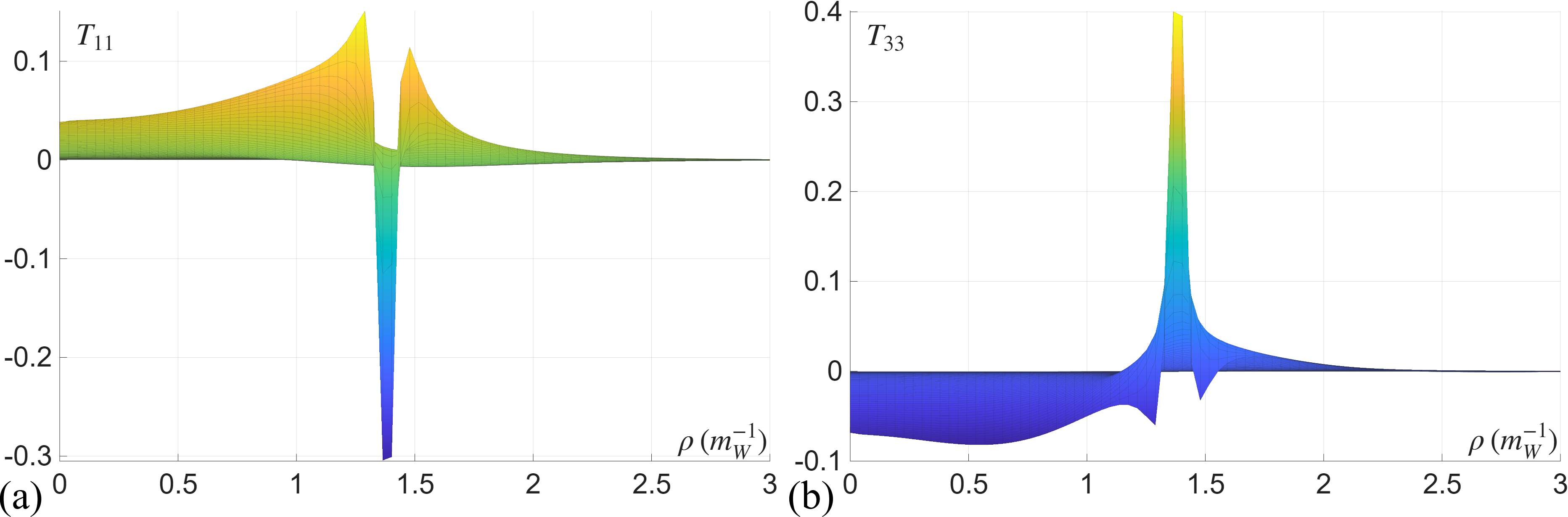}
		\caption{$T_{ii}$--$\rho$ views of the spatial distribution for stress-energy components (a) $T_{11}$ and (b) $T_{33}$ for $n=3$ electroweak vortex ring solutions with physical parameters $\beta=0.7782$ and $\theta_W=28.18^\circ$.}
		\label{fig:T11T33}
	\end{figure}

	Figure~\ref{fig:T11T33} displays the spatial distributions of stress-energy components $T_{ii}$ for the $n=3$ electroweak vortex ring with physical parameters.
	In both panels, a spike is observed at precisely the vortex ring location ($H=0$). These features arise because the stress-energy components $T_{ii}$ involve derivatives of profile functions $\Phi_1$ and $\Phi_2$. The magnitudes of these spikes remain moderate---approximately four times higher than the surrounding field values---ensuring that the neighboring physical structures are fully resolved. Rather than representing physical singularities, these spikes are numerical artifacts originating from the steep gradients in the Higgs sector. To verify this, we performed convergence tests across three distinct grid choices ($70 \times 60$, $70 \times 72$, and $160 \times 60$) and observed a consistent reduction in spike amplitude, confirming that these features can be smoothed out in the continuum limit.
	
	An intriguing feature of the stress distribution, as seen in figure~\ref{fig:T11T33}, is that the spatial profiles of $T_{11}$ and $T_{33}$ act effectively as mirror images of one another. This reciprocal relationship can be understood through a mechanical analogy to an incompressible material, such as a pizza dough: a compression in one principal direction necessitates a transverse expansion in the orthogonal direction.
	
	In a perfectly symmetric setting, the Bennett pinch typically manifests in the stress-energy tensor components as a consistent negative (attractive) value, representing the self-stabilization of the plasma. However, the electroweak vortex rings possess a more complicated internal structure. As demonstrated in figure~\ref{fig:currents}(a), the neutral currents $j_i^Z$ form double concentric, counter-rotating loops, with the outer ring carrying a higher current density. This internal architecture results in the pinch effect having both attractive and repulsive components. For a massive field like the $Z$-boson, the Bennett effect associated with it is expected to be intense and short-ranged---a profile that these localized spikes mirror with remarkable precision.
	
	The presence of such a pinch is the logical corollary to the neutral Amp\`ere's law observed in our solutions: if the $j_i^Z$ currents generate a concentric neutral field, then the reciprocal back-action of this field upon the currents is physically requisite, manifesting as the self-constricting pressure of the Bennett effect. While this numerical evidence is highly promising, the definitive isolation of the neutral Bennett pinch remains an open challenge. Future investigations employing finer mesh densities and Abelian decomposition~\cite{AD1,AD2,AD3,AD4,AD5,AD6} will be essential to decouple the $Z$-boson's specific contributions from the composite stress-energy components. Such refinements may ultimately yield an explicit analytical scaling law, directly correlating the $Z$-boson mass with the magnitude and spatial confinement of the neutral Bennett pinch.

	\section{Conclusion}
	We have demonstrated that the stabilization paradigm identified in Cho-Maison MAPs extends naturally to the electroweak vortex rings. These configurations carry no topological monopole charge, confirming that the repulsive interactions are intrinsic and fundamental to the Weinberg-Salam model. The interplay of these repulsions fully determines the geometric size of the vortex rings, quantified by the radius $R_\rho$, and governs their total energy $E$. For physical model parameters, the solutions possess masses of $18.01$~TeV ($n=2$) and $26.80$~TeV ($n=3$).
	
	The field distribution within these configurations exhibits an intricate spatial pattern. Specifically, the charged $W$-boson currents form a toroidal-poloidal Hopf-type structure, with a $\pi/2$ phase relation between their individual components and an azimuthally alternating flow pattern characterized by $2n$ lobes. In contrast, the electromagnetic and $Z$-boson currents form two concentric, counter-rotating loops, consistent with the classical Amp\`ere's circuital law and its neutral analogue discussed above. The structural similarities between electroweak theory and condensed matter systems may also provide motivation for exploring analogous current-field relations in multi-component superfluids and superconductors.
	
	Stress tensor analysis uncovers pronounced oscillations in both horizontal and vertical directions, showing that variations in the Higgs sector subtly modulate the stability of the configuration. These non-monotonic transitions reflect a competition between bosonic repulsive interactions and the Bennett pinch. This interplay suggests that the electroweak vortex rings possess an inherent propensity for instability. Such oscillatory behaviour could fundamentally influence the decay channels of these structures, potentially leaving distinct signatures from the electroweak epoch that warrant further investigation in future high-energy experiments.
	
	Finally, our mass predictions place the lightest configuration ($n=1$, $E<14$ TeV) theoretically within the reach of the LHC, albeit with a minuscule cross-section. Extrapolating from the discovery of the 125.1 GeV Higgs boson at $\sqrt{s}=7$--$8$ TeV~\cite{HiggsAtlas,HiggsCMS}, a conclusive observation would likely require the proposed FCC-hh. Beyond their mass, these structures are shown to be unstable saddle-point solutions characterized by a baryon number $Q_B=n/2$~\cite{wsvortex1,wsvortex2,Teh}. Nonetheless, the correspondence with superconductors suggests that meta-stability could potentially be achieved in the dense, high-temperature plasma of the early universe. In this context, these objects represent energy barriers in the Standard Model vacuum. Although their contribution to electroweak baryogenesis is expected to be subdominant due to stronger Boltzmann suppression associated with their significantly higher mass, further study of these solutions would enrich our understanding of the spectrum of saddle-point configurations relevant to baryon number violation.

	\acknowledgments
	This work has been supported by the Research Program of State Key Laboratory of Heavy Ion Science and	Technology, Institute of Modern Physics, Chinese Academy of Sciences (Grant No. HIST2025CS08), and the National Key R\&D Program of China (Grant
	No. 2024YFE0109800 and 2024YFE0109802). The authors thank Prof. Pengming Zhang and Dr. Liping Zou for useful discussions and comments.

	\bibliographystyle{JHEP}
	\bibliography{references.bib}
\end{document}